\documentclass[lettersize,journal, openany]{IEEEtran}

\usepackage{booktabs}
\usepackage{amssymb}
\usepackage{algorithmic}
\usepackage{algorithm}
\usepackage{array}
\usepackage{textcomp}
\usepackage{url}

\usepackage{verbatim}
\usepackage{graphicx}
\usepackage{cite}
\usepackage{relsize}
\usepackage{etoolbox}
\usepackage{flushend}
\usepackage{balance} 
\usepackage{cuted}     
\usepackage{amsthm}
\newtheoremstyle{case}{}{}{}{}{}{:}{ }{}
\usepackage{mathtools}
\usepackage[margin = 0.9in]{geometry}
\usepackage{diagbox}
\usepackage{multirow}
\usepackage{blindtext}
\usepackage{comment}
\usepackage{caption}
\usepackage{hyperref}

\makeatletter
\patchcmd{\@makecaption}
  {\scshape}
  {}
  {}
  {}
\makeatother
\usepackage[section]{placeins}
\usepackage{placeins}
\ifCLASSOPTIONcompsoc
    \usepackage[caption=false, font=normalsize, labelfont=sf, textfont=sf]{subfig}
\else
\usepackage[caption=false, font=footnotesize]{subfig}
\fi
\begin{document}

\title{Online High-Frequency Trading Stock Forecasting with Automated Feature Clustering and Radial Basis Function Neural Networks\textsuperscript{\dag}\thanks{\dag This paper was presented at the Economics of Financial Technology Conference, held from 21\textsuperscript{st} to 23\textsuperscript{rd} June 2023, in Edinburgh, UK.}}

\author{\IEEEauthorblockN{Adamantios Ntakaris\IEEEauthorrefmark{1}, Gbenga Ibikunle\IEEEauthorrefmark{1},  
}\\
\IEEEauthorblockA{\IEEEauthorrefmark{1}Business School,
University of Edinburgh, Edinburgh, EH8 9JS, UK}\\
}

\maketitle
\begin{abstract}
This study presents an autonomous experimental machine learning protocol for  high-frequency trading (HFT) stock price forecasting that involves a dual competitive feature importance mechanism and clustering via shallow neural network topology for fast training. By incorporating the k-means algorithm into the radial basis function neural network (RBFNN), the proposed method addresses the challenges of manual clustering and the reliance on potentially uninformative features. More specifically, our approach involves a dual competitive mechanism for feature importance, combining the mean-decrease impurity (MDI) method and a gradient descent (GD) based feature importance mechanism. This approach, tested on HFT Level 1 order book data for 20 S\&P 500 stocks, enhances the forecasting ability of the RBFNN regressor. Our findings suggest that an autonomous approach to feature selection and clustering is crucial, as each stock requires a different input feature space. Overall, by automating the feature selection and clustering processes, we remove the need for manual topological grid search and provide a more efficient way to predict LOB's mid-price.
\end{abstract}

\begin{IEEEkeywords}
High-frequency trading, limit order book, online learning, stock forecasting, clustering, neural networks
\end{IEEEkeywords}

\section{Introduction}
\IEEEPARstart{F}{inancial} markets are constantly changing, with the rise of technology creating an environment where transactions can occur in fractions of a second. High-frequency trading (HFT) plays a critical role in this context, leveraging advanced algorithms to execute a large number of orders at very fast speeds. HFT, driven by machine learning (ML) algorithms, has introduced a dynamic and extremely fast-paced level of complexity into that trading universe. However, these ML algorithms often rely on predictive experimental protocols that use potentially uninformative and noisy features. These features are selected based on computationally expensive optimization routines and manual methods, heavily dependent on the trader's domain knowledge for the forecasting task. This manual approach of operation presents significant challenges in the HFT forecasting environment, where the speed of decision-making is critical. A fully autonomous feature importance and feature clustering routine as part of predictive machine learning experimental protocol presents a promising solution to these challenges. 

A typical machine learning experimental protocol contains several parts such as data processing, feature extraction, feature importance (and/or feature selection), generate input matrix, input matrix clustering, and model selection (i.e., regressor, classifier or both). Each of these parts require extensive domain knowledge and several manual actions such as feature importance and input clustering (i.e., number of clusters). In this study, we focus on the automation of these two parts (i.e., feature importance and input clustering) to create optimized, responsive and online trading routines in the HFT universe. We do that by utilizing the MDI and GD methods as feature importance mechanisms. We selected GD as a benchmark method to MDI. Then, MDI and GD separately and combined with the features' correlation matrix will guide the k-means clustering method to the RBFNN regressor that will predict LOB's mid-price. 

MDI \cite{randforest}, \cite{gini}, also known as Gini importance, is a measure of the total reduction in the impurity brought about by a feature in a tree-based model, averaged across all the trees in the model. The effectiveness of that method was tested successfully on problems such as digit recognition problems \cite{louppe}, \cite{sutera}, predict functional labels associated with genomic regions in the ChIP dataset \cite{Li}, and predicted daily global solar radiation \cite{solar}. This method has also been utilized in finance \cite{sita}, \cite{magesh}, \cite{lopez}. GD is also a method that has been extensively employed since its inception \cite{cauchy} in several different fields such as the development of linear adaptive filters \cite{filter}, for the computation of the Hessian vector products \cite{hessian}, and the approximation of large matrices from the Netflix competition \cite{netflix}. The additional components of this fully autonomous process, the k-means clustering and the RBFNN regressor, have also demonstrated their capabilities in various tasks. More specifically, k-means algorithm effectively tested on image classification and segmentation datasets \cite{likas}. RBFNN proposed as an effective classifier for tasks derived from the UCI machine learning repository for image segmentation \cite{sure}. Both methods have been also employed for financial tasks such as stock price prediction and \cite{dash}, \cite{xu}. 

Despite the extensive range of applications of the aforementioned methods, their implementation has been static during the training process, which means that the feature importance and the number of clusters within the k-means algorithm had to be selected manually or based on the elbow method - a heuristic method that determines the number of clusters by plotting the explained variation as a function of the number of clusters. To the best of our knowledge, this is the first time that an online and fully autonomous feature importance and optimized identification of the number of clusters in the HFT domain is presented in the literature. 

The rest of this paper is organized as follows. In \hyperref[section:Related]{Section \ref{section:Related}}, we provide the related literature review to the previous mentioned methods. In \hyperref[section:Method]{Section \ref{section:Method}} we provide the technical details of our fully autonomous pipeline and in \hyperref[section:Expe]{Section \ref{section:Expe}} we present the results of our experimental and comparative analysis and also discuss limitations and future research directions. In \hyperref[section:Con]{Section \ref{section:Con}}, we summarize our findings. 

\section{Related Work}\label{section:Related}

\noindent The implementation of a fully automated machine learning experimental protocol, similar to this study, consists of several parts. More specifically, the related parts are feature extraction, feature importance (and/or feature selection), potential clusters that may exist among the extracted/selected features, and the type of forecaster. Several studies have tried to provide an automated framework for feature selection. For instance, Authors in \cite{penn} utilized reinforcement learning methods to automate the trading process tested on the Penn Exchange Simulator (PXS), but only achieved reasonable heuristic marginal gains. Authors in \cite{medi} developed an automated feature selection of predictors for electronic medical records data based on a two-step process involving clustering and regularized regression with the use of a parametric mixture model. There are also additional studies that performed automated feature selection for financial forecasting tasks \cite{zhang, hejek}. 

Regarding the specific methods that we utilize for feature importance and clustering, MDI has proven to be an effective method for feature selection in water absorption in oilfields in China \cite{liu}, and factors affecting the willingness to pay (WTP) in the agricultural insurance space \cite{anisa}. MDI has also been utilized on financial data forecasting tasks such as the prediction of clean energy US ETFs \cite{sado} and factors that can affect stock prices \cite{prado}. We also employ GD in our forecasting experimental protocol. GD is one of the core optimization methods in the machine learning literature. GD examined for its effectiveness as part of fixed point and trajectory learning for discrete and continuous models under forward and recurrent architectures \cite{baldi}. GD also deployed to facilitate the development of new reinforcement learning (RL) algorithms \cite{baird}.

Our fully autonomous framework is based also on two widely used machine learning methods such as the k-means clustering algorithm and RBFNN. The k-means algorithm and its improved variants have been analyzed based on their theoretical properties around the concepts of initialization (i.e., starting points/values) and methods to define the optimal number of clusters \cite{kmeans}. K-means algorithm was extensively used in finance for several different clustering tasks. For instance, authors in \cite{indian}, used the algorithm to find cluster of stocks for optimal returns in the Indian stock market. The k-means algorithm has been effectively combined with the RBFNN. More specifically, authors in \cite{oh} combined effectively k-means algorithm with RBFNN forecasting tasks based on the synthetic datasets and two widely used publicly available datasets such as the Automobile Miles per Gallon (MPG) and Boston Housing (BH) dataset. The combination of these two algorithms was also utilized for financial forecasting tasks. Fro example, authors in \cite{shen} utilized an optimized version of k-means based on artificial fish swarm algorithm (AFSA) and combined it with RBFNN for forecasting stock indices of the Shanghai Stock Exchange. Their implementation has also been tested efficiently for the prediction of the LOB's mid-price by the authors in \cite{vivi}. 

Despite the extensive and effective use of the methods above the HFT universe requires fully automated processes that will reduce the decision making time in an online manner. To the best of our knowledge, this is the first study on automated feature importance and input clustering in the HFT and LOB literature.   

\section{Proposed Method}\label{section:Method}

\noindent The rate of information flow in the HFT LOB  universe requires agile methods for fast decision execution. We tackle this challenge by developing a fully automated mechanism with respect to the feature importance (i.e., information importance) and the optimal number of information clusters as part of a machine learning experimental protocol. This protocol is compartmentalized into the following four blocks:

\begin{itemize}
\item Block 1: Feature importance competitive mechanism based on MDI and GD,
\item Block 2: Construct the correlation-based observation matrix,
\item Block 3: Define the optimal number of clusters via the k-means algorithm and the silhouette scores,
\item Block 4: Employ the RBFNN regressor.
\end{itemize}

More specifically, Block 1 is based on the development of a feature importance mechanism that will highlight which features are more relevant/important for the task of mid-price forecasting - a forecasting task that considers the average price of the best bid and the best ask LOB prices. Before we delve into the details of the MDI and GD methods we should mention that the primal objective of this Block (i.e., Block 1) is to highlight the utility of the MDI method. To achieve this we need to establish a benchmark method such as the GD algorithm - a basic machine learning optimization algorithm. 
GD is commonly utilized as an optimization and iterative method, and we will convert it to a feature importance technique. Both methods will operate under an online learning experimental protocol. This means that the forecasting results will be reported per trading event arrival (i.e., tick-by-tick data) and will not be affected by the time of arrival.

MDI is a technique that assesses feature importance, especially in ensemble models such as the random forest (RF). Specifically, in the case of a RF regression, MDI provides a quantitative measure of the contribution of individual features to the forecasting ability of the regressor, which is based on the decrease in node impurity averaged over all trees in RF. The impurity of a tree node is measured by the mean squared error (MSE) of the target variable. The variance for a node $j$ of the feature $f$ is given by:

\begin{equation}
I_{j,f} = MSE_{j,f} = \frac{1}{N_j} \sum_{i \in D_j}^{}(y_i - \tilde{y}_j)^2,
\end{equation}
\noindent where $N_j$ is the number of samples at node $j$, $D_j$ is the set of training samples per node $j$, $y_i$ is the target value for sample $i$, and $\tilde{y}_t$ is the average target value for the samples at node $j$. The impurity reduction $\Delta I$ per node $j$ of the feature $f$ is calculated, as follows:

\begin{equation}
\Delta I_{j,f} = I_{j,f} - \Big ( \frac{N_l}{N_j}\Big )I_l - \Big ( \frac{N_r}{N_j}\Big )I_r, 
\end{equation}

\noindent where:
\begin{itemize}
\item $N_l$ and $N_r$ are the numbers of samples that go to the left and right child nodes, respectively, after the split, 
\item $I_l$ and $I_r$ are the impurities of the left and right child nodes, respectively, after the split,
\item and $N_j$ is the total number of samples at the current node. 
\end{itemize}

\noindent then the MDI per feature $f$ is calculated, as follows:

\begin{equation}\label{eq:mdif}
mdi_f = \frac{1}{B} \sum_{b=1}^{B} \Big (\sum_{j \in P(b,f)}^{}\Delta I_{j,f}\Big ),
\end{equation}

\noindent where $B$ is the total number of tress in RF, $j$ is the node that splits  on feature $f$, $P(b,f)$ is the set of nodes that split on feature $f$ in tree $b$, and $mdi_f \in {\mathbb{R}^{F}}$ for $F$ features. 

The second method that we employ in Block 1 is the GD method. We convert GD to a feature importance method following the implementation of authors in \cite{optm}. More specifically, the observation matrix $X_t \in \mathbb{R}^{N \times F}$, where $t$ represents the current state of the input information of size $N$, is attached to the weight vector $\theta$. The weight vector $\theta$ will be updated iteratively based on GD following the algorithmic process in \hyperref[algo:gd]{Algorithm \ref{algo:gd}}.

\begin{algorithm}
  \caption{Gradient Descent Feature Importance}
  \begin{algorithmic}[1]
    \REQUIRE Learning rate $\alpha$, number of iterations $R$, $\theta$, labels $y_{t}$
    \FOR{$i=1$ to $R$}
      \STATE Predicted labels: $\hat{y_t}$ = $\sum_{i=1}^{N}\theta^T \cdot X_t$
      \STATE $error$ = $\hat{y_t} - y_{t}$
      \STATE Gradient: $\nabla J(\theta)$ = $2 \cdot (X_t^T \cdot error)$
      \STATE Parameter update: $\theta \gets \theta - \alpha \cdot \nabla J(\theta)$
    \ENDFOR
    \ENSURE Optimized vector $\theta$.
  \end{algorithmic}
  \label{algo:gd}
\end{algorithm}

The optimized vector $\theta \in \mathbb{R}^F$ operates now as the trained importance weight vector. Both vectors, $\theta$ and $mdi_f$ will then be attached via a matrix-vector product (i.e., broadcasting) to the observation matrix:

\begin{equation}\label{eq:fimdi}
FI_{MDI} = X_{if} \cdot mdi_f, \ 1 \leq i \leq N, \ 1 \leq f \leq F,
\end{equation}

\begin{equation}\label{eq:figd}
FI_{GD} = X_{if} \cdot gd_f, \ 1 \leq i \leq N, \ 1 \leq f \leq F,
\end{equation}

\noindent where $i, f \in \mathbb{N}^+$, and $mdi_f$ and $gd_f$ are the competitive feature importance vectors based on MDI (i.e., \hyperref[eq:mdif]{Eq. \ref{eq:mdif}} ) and GD (i.e., $\theta$ vector in \hyperref[algo:gd]{Algorithm \ref{algo:gd}}) methods, respectively.

The next step in the experimental pipeline is Block 2, where we express the two competitive matrices $FI_{MDI}$ and $FI_{GD}$ as correlation matrices. These two correlation matrices then will be converted to distance matrices following the authors in \cite{lewis}, as described below:

\begin{equation}\label{eq:dist}
C = \sqrt{\frac{1}{2}(1 - \rho_{m,n})},
\end{equation}

\noindent where $\rho_{m,n}$ represents the MDI correlation-based matrix and is defined, as follows:

\begin{equation}\label{eq:corr}
\rho_{m,n} = \frac{\sum\limits_{i=1}^{N} (\textbf{x}_{i,m} - \bar{\textbf{x}}_m) (\textbf{x}_{i,n} - \bar{\textbf{x}}_n)}{\sqrt{\sum\limits_{i=1}^{N} (\textbf{x}_{i,m} - \bar{\textbf{x}}_m)^2 \sum\limits_{i=1}^{N} (\textbf{x}_{i,n} - \bar{\textbf{x}}_n)^2}},
\end{equation}

\noindent where $\textbf{x}_{i,m}$ denotes the $m^{th}$ feature (column) of the $i^{th}$ data point, $\textbf{x}_{i,n}$ denotes the $n^{th}$ feature (column) of the $i^{th}$ data point, and finally $\bar{\textbf{x}}_m$ and $\bar{\textbf{x}}_n$ are the means of the $m^{th}$ and $n^{th}$ feature sets over all data points, respectively.

The same equations (i.e., \hyperref[eq:dist]{Eq. \ref{eq:dist}} and \hyperref[eq:corr]{Eq. \ref{eq:corr}}) express the case of the GD-based correlation matrix. Motivation for expressing the input information as a correlation-based distance matrix (see \hyperref[eq:dist]{Eq. \ref{eq:dist}}) is the fact that correlation in high-dimensional datasets such as HFT LOB datasets can increase the interpretability of the raw data. For instance, if two variables are highly correlated then they will have a low distance in the matrix (see \hyperref[eq:dist]{Eq. \ref{eq:dist}}). An additional advantage of this transformation is that certain algorithms such as k-means require a measure of similarity and distance calculation between data points.  

The next phase is the development of Block 3 that will define the optimal number of clusters under the k-means clustering method by calculating the silhouette score, a score that considers both the tightness of the clusters and the separation between them. Following \cite{lewis}, the calculation of the silhouette score is a two-step process and it based on the silhouette coefficients $S_i$ and a quality ratio $q$ that considers the mean and variance of the silhouette coefficients, as follows:

\begin{equation}
S_i = \frac{b_i - a_i}{max\{a_i,b_i\}}, 
\end{equation}

\begin{equation}
q = \frac{E[\{S_i\}]}{\sqrt{V[\{S_i\}]}},
\end{equation}
\noindent where $ 1 \leq i \leq N$, $i \in \mathbb{N}^+$, $a_i$ is the average distance between $i$ and the rest of the elements within the same
cluster and $b_i$ is the average distance between $i$ and the rest of the elements in the nearest cluster that $i$ is not a member. As a result a higher $q$ value is  preferable. 

The silhouette and quality scores are part of Block 4, the final part of the experimental protocol which is an iterative algorithm that considers/fit the k-means clustering to the observation matrix (i.e., the distance-based correlation matrix in \hyperref[eq:dist]{Eq. \ref{eq:dist}} or $C = \{\textbf{c}_1, \textbf{c}_2, ... \textbf{c}_F\}$). The main objective of the k-means algorithm is to partition the observation matrix $C$ into $K$ groups $G = \{ G_1, G_2, ..., G_K\}$ so as to minimize the within-cluster sum of squares, via a three-step iterative process, where the main objective is: 

\begin{equation}
\underset{G}{\arg\min} \sum_{k=1}^{K} \sum_{\textbf{c} \in G_k} ||\textbf{c} - \mu_k||^2
\end{equation}

\noindent where $\textbf{c}$ represents the points that belong to cluster $G_k$ and the steps are:

\begin{itemize}
\item Initialize randomly the centroids/clusters
\begin{equation}
\mu_k^{(0)}, \text{ for } \ 1 \leq i \leq K \text{ and } k, \ K \in \mathbb{N}^+  
\end{equation}
\item Assign (i.e., euclidean distance-based assignment) each point to the nearest clusters, according to:
\begin{equation}
G_k^{(t)} = \{c_p: ||c_p - \mu_k^{(t)}||^2 \leq ||c_p - \mu_j^{(t)}||^2, 1 \leq j \leq K \}
\end{equation}
where $t$ is the iteration index and $c_p$ is the $p^{th}$ data point,
\item Centroids update:
\begin{equation}
\mu_k^{(t+1)} = \frac{1}{|G_k^{(t)}|} \sum_{c_p \in G_k^{(t)}} c_p
\end{equation}
\end{itemize}

\noindent The extraction of the updated centroids (i.e., convergence of the k-means algorithm) is the final step of Block 4. We can summarize the Block 4 in the following algorithm \hyperref[algo:clustering]{Algorithm \ref{algo:clustering}}:

\begin{algorithm}
  \caption{Clustering Quality}
  \begin{algorithmic}[1]
    \REQUIRE Dataset $C$, Maximum number of clusters $maxClusters$, Number of iterations $n\_iter$
    \FOR{$init=1$ to $n\_init$}
      \FOR{$num\_clusters=2$ to $maxClusters$}
        \STATE Initialize k-means with $num\_clusters$
        \STATE Fit k-means to $C$
        \STATE Compute $silh$ for each sample in $C$
        \STATE Calculate $mean\_silh$ and $var\_silh$ 
        \STATE Compute $q = mean\_silh / var\_silh$
        \IF{($q$ is higher than the best ratio)}
          \STATE Update the best $q$
          \STATE Update the best $num\_clusters$
          \STATE Save current silhouette scores
        \ENDIF
      \ENDFOR
    \ENDFOR
    \STATE Reorder the correlation matrix according to the best clustering arrangement
    \ENSURE Best number of clusters.
  \end{algorithmic}
  \label{algo:clustering}
\end{algorithm}

The final part of the fully autonomous experimental pipeline is the implementation of the RBFNN regressor that consists of the input layer, the hidden layer and the output layer. More specifically, the inputs to the RBFNN are $FI_{MDI}$ and $FI_{GD}$ (see \hyperref[eq:fimdi]{Eq. \ref{eq:fimdi}} and \hyperref[eq:figd]{Eq. \ref{eq:figd}}), the number of centroids is determined by the silhouette scores and the clustering quality measure $q$, as these described in \hyperref[algo:clustering]{Algorithm \ref{algo:clustering}}. Next, the hidden layer contains the radial basis function (RBF) activation function:

\begin{equation}\label{eq:rbf}
RBF(x_i, \mu_j, \sigma_j) = \alpha_{i,j} = e^{-\frac{{\lVert x_i - \mu_j \rVert}^2}{{2\sigma_j^2}}}
\end{equation}

\noindent where $x_i$ is the $i^{th}$ transformed data sample (i.e., based on $FI_{MDI}$ or $FI_{GD}$ input transformations), $\mu_j$ and $\sigma_j$ represent the center and standard deviation of the $j^{th}$ RBF neuron, respectively, as follows:

\begin{equation}
\sigma_j = \frac{1}{{\text{{K}}(\text{{K}} - 1)}} \sum_{l=1}^{\text{{K}}} \sum_{m \neq l}^{\text{{K}}} \lVert \mu_l - \mu_m \rVert.
\end{equation}

\noindent where the expression of the spread is a heuristic approach that considers the comparison of the center $l$ with every other center while calculating the sum of distances from center $l$ to all other centers. After the calculation of the RBF-based activation functions we proceed to the output layer where the prediction of the labels (i.e., regression) is based on the collection/summation of the entire line-up of trained weights and activation results, as follows:

\begin{equation}
\tilde{y}_i= \sum_{k}^{K}\alpha_{i,k} \cdot w_{k},
\end{equation}

\noindent where $w_j$ represents the weights between the hidden and the output layer for the $j^{th}$ RBF neuron and is calculated, as follows:

\begin{equation}
\mathbf{w} = (\mathbf{A}^\top \mathbf{A})^{-1} \mathbf{A}^\top \mathbf{y},
\end{equation}

\noindent where $\mathbf{w}$ represents the weights, $\mathbf{A}$ is the RBF activation matrix of the hidden layer, $\mathbf{A}^\top$ denotes the transpose of matrix $\mathbf{A}$, $\mathbf{y}$ is the target vector, and $(\mathbf{A}^\top \mathbf{A})^{-1}$ represents the inverse of the matrix product $\mathbf{A}^\top \mathbf{A}$.

\section{Experiments}\label{section:Expe}

\noindent LOB is an important tool for the HFT trading information flow analysis. More specifically, LOB is defined as a price and volume classification mechanism that the HFT ML trader conditioning on the state of the market will decide to place an order to buy, sell (or cancel) a stock. This trading activity creates an extremely dynamic environment around the best prices within LOB. The best LOB price level contains the best ask and the best bid prices, with their average being referred to as the mid-price - a good proxy that can act as a sensitive indicator of the LOB state and it is defined in the following way:

\begin{equation}
MP_t = \frac{P_{A_t} + P_{B_t}}{2},
\end{equation}

\noindent where $P_{A_t}$ and $P_{B_t}$ are the best ask and the best bid prices at trading event $t$, respectively.

\subsection{Experimental Protocol and Dataset}
\noindent The main objective of our experimental protocol is to predict the mid-price (i.e., regression task) in an online manner (i.e., tick-by-tick forecasting) via a fully autonomous process with respect to feature importance and number of clusters. In \hyperref[section:Method]{Section \ref{section:Method}} we provided the proposed mechanics of the fully autonomous framework. A higher overview of that process is that we utilize two competitive methods (i.e., MDI and GD) as feature importance mechanisms that will be attached to the same input matrix and then via a distance-based correlation transformation, the input will be then fed to RBFNN. More specifically, RBFNN will operate using the k-means algorithm, which will provide the RBF neuron with the centroids and their equivalent standard deviations via an autonomous process that relies on the silhouette score and a clustering quality indicator that defines the number of centroids within RBFNN. Motivation for the development of this fully autonomous and comparative study is to highlight that forecasting tasks in the HFT universe requires online and agile concepts that do not rely on stale training processes. 

\begin{figure*}[!htpb]
    \centering
    \includegraphics[width=14cm]{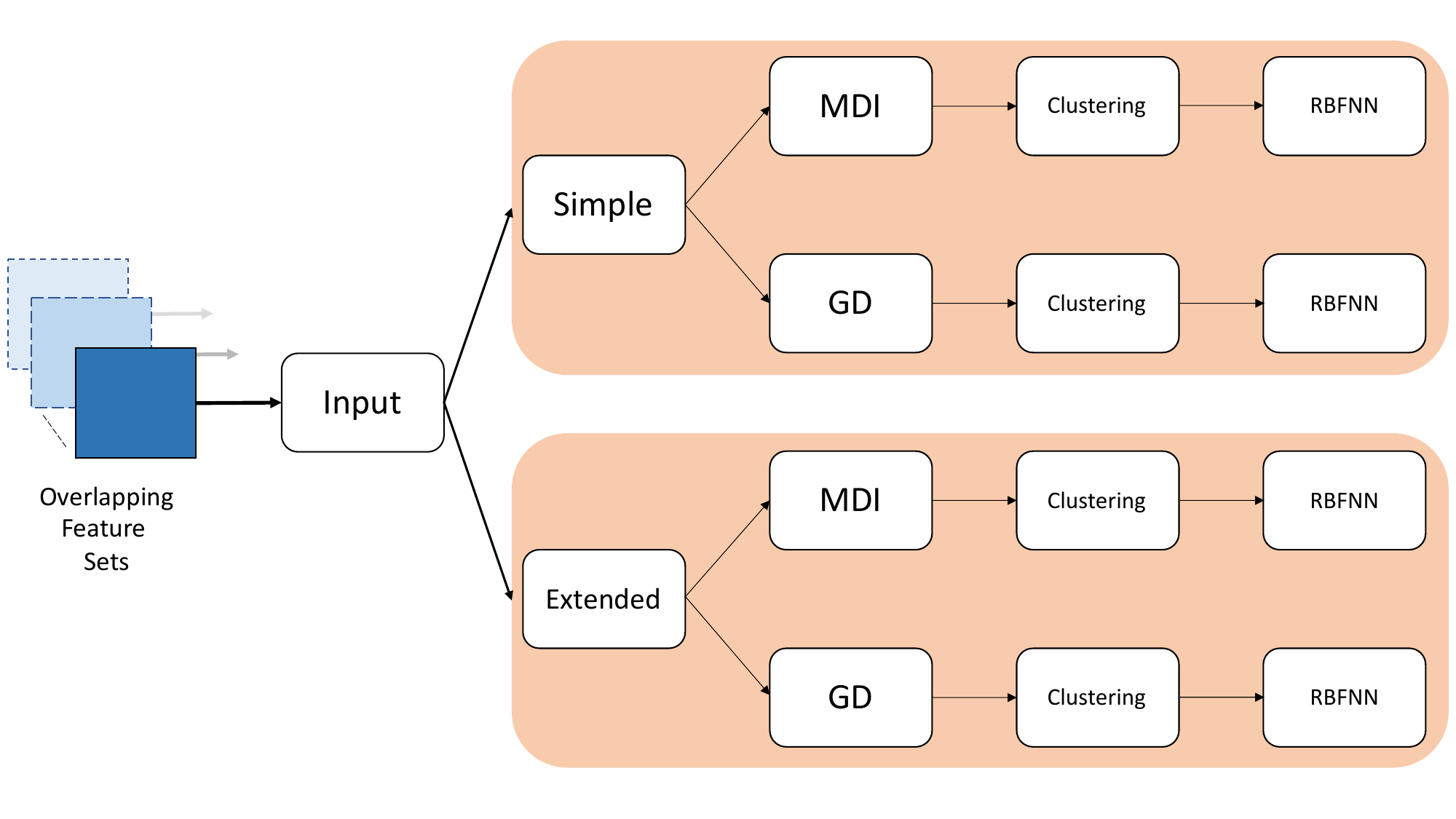}
    \caption{Overview of the fully automated protocol. From left to right: The first part of the online experimental protocol is the transformation of the LOB data to sliding window data blocks. Each of these data blocks are fed sequentially to the fully automated mechanism. Within that mechanism we have two competitive pipelines (i.e., MDI and GD) that will provide the feature importance vectors. Then the clustering block defines the optimal number of clusters based on the weighted by the feature importance vectors input matrix (i.e., correlation distance-based matrix). The clusters then determine the centroids and the standard deviation of the RBF neurons. The number of clusters is changing constantly (i.e., online) based on the latest input feature set.}
    \label{fig:protocol}
\end{figure*}

The experimental protocol is also equipped with two feature sets that will help us to understand how different information can affect not only the mid-price forecasting but also the alternation frequency between the two feature importance methods and the number of clusters per trading event. More specifically, the two features sets have the following characteristics: the first set, named Simple in our experimental protocol, consists of the best ask and the best bid prices together with their equivalent trading volumes and can be found in \hyperref[tab:feat]{Table \ref{tab:feat}} as LOB Best Level. The second feature set, named as Extended in our experimental protocol, is based on a collection of basic, kernelized and polynomial features and their description can be found in \hyperref[tab:feat]{Table \ref{tab:feat}} as Basic, Synthesized, Linear Kernel, Polynomial Kernel, Sigmoid Kernel, Experimental Kernel, and RBF Kernel. Motivation for the selected features is to expand the Simple feature set to linear and non-linear transformations of LOB's best ask and best bid prices (and their equivalent volumes). An overview of the online experimental protocol can be seen in \hyperref[fig:protocol]{Fig. \ref{fig:protocol}}. The size of the sliding window blocks is set to 100 events per block with an overlap of 99 events compared to the previous block. We noticed that a size of 100 is sufficient for clustering and time efficiency in terms of forecasting speed. We should also mention that training and testing splits are based on a cumulative five-fold setting which means that every training fold absorbs the available testing set and converts it to training data after the prediction performance is reported. This way the training set acts as a cumulative set that sequentially stores the latest HFT data inflow. The performance reporting is based on Mean Squared Error (MSE) and Root Mean Squared Error (RMSE), which are calculated as follows:

\begin{equation}
MSE = \frac{1}{N} \sum_{i=1}^{N} (y_i - \hat{y}_i)^2
\end{equation}
and 

\begin{equation}
RMSE = \sqrt{\frac{1}{N} \sum_{i=1}^{N} (y_i - \hat{y}_i)^2}
\end{equation}

\noindent where $N$ is the total number of samples, $y_i$ are the actual values, and $\hat{y}_i$ are the predicted values. Motivation for using these two reporting metrics is the fact that both have the ability to highlight large errors and also RMSE produces results in units that are similar to the predicted task.

\begin{table}[!hbtp]
\centering
\captionsetup{width=12.00\textwidth}
\caption{Feature Sets}
\scalebox{1.10}{
\begin{tabular}{r l}
\toprule
\textbf{Set} & \textbf{Feature} \\
\hline
LOB Best Level & $u_1=\{ P^{ask}_1, V^{ask}_1, P^{bid}_1, V^{bid}_1\}$ \\
Basic & $u_2=P^{ask}_1 + P^{bid}_1$ \\
 & $u_3=P^{ask}_1 - P^{bid}_1$ \\
 & $u_4=\sin(P^{ask}_1 P^{bid}_1)$ \\
Synthesized & $u_5=P^{ask}_1 P^{bid}_1$ \\
 & $u_6=V^{ask}_1 V^{bid}_1$ \\
 & $u_7={P^{ask}_1}^2 + {P^{bid}_1}^2$ \\
 & $u_8={V^{ask}_1}^2 + {V^{bid}_1}^2$ \\
Linear Kernel & $u_9=P^{ask}_1 P^{bid}_1$ \\
Polynomial Kernel & $u_{10}=(P^{ask}_1 P^{bid}_1 + c_0)^d$ \\
Sigmoid Kernel & $u_{11}=\tanh(\gamma P^{ask}_1 P^{bid}_1 + c_0)$ \\
Exponential Kernel & $u_{12}=e^{-\gamma |P^{ask}_1 - P^{bid}_1|}$ \\
RBF Kernel & $u_{13}=e^{-\gamma (P^{ask}_1 - P^{bid}_1)^2}$ \\
\bottomrule
\end{tabular}}
\label{tab:feat}
\end{table}

The HFT data in the present study is based on tick-by-tick LOB Level 1 data from 20 US Nasdaq- and NYSE-listed mega cap (i.e., companies with market value of \$200 billion or more) companies in a nanosecond time resolution from Refinitiv. The list of companies  can be found in \hyperref[tab:company]{Table \ref{tab:company}} and the covered trading period is three months from $1^{st}$ of September 2022 to $30^{th}$ of November 2022.  

\begin{table}[!hbtp]
\centering
\captionsetup{width=12.00\textwidth}
\caption{Company Names and Ticker Symbols}
\scalebox{1.00}{
\begin{tabular}{r l}
\toprule
\textbf{Ticker Symbol} & \textbf{Company Name} \\
\hline
AMZN & AMAZON.COM, INC. \\
BAC & BANK OF AMERICA CORPORATION \\
BRK & BERKSHIRE HATHAWAY INC. \\
GOOGL & ALPHABET INC. \\
JNJ & JOHNSON \& JOHNSON \\
JPM & JPMORGAN CHASE \& CO. \\
KO & THE COCA-COLA COMPANY \\
LLY & ELI LILLY AND COMPANY \\
META & META PLATFORMS INC. \\
MRK & MERCK \& CO., INC. \\
MSFT & MICROSOFT CORPORATION \\
NVDA & NVIDIA CORPORATION \\
NVO & NOVO NORDISK A/S \\
ORCL & ORACLE CORPORATION \\
PEP & PEPSICO, INC. \\
PG & THE PROCTER \& GAMBLE COMPANY \\
UNH & UNITEDHEALTH GROUP INCORPORATED \\
VISA & VISA INC. \\
WMT & WALMART INC. \\
HD & THE HOME DEPOT, INC. \\
\bottomrule
\end{tabular}}
\label{tab:company}
\end{table}

\subsection{Results}

\noindent The objective of the present study is to predict LOB's mid-price via a fully automated process with respect to feature importance and number of clusters. The experimental protocol follows the online forecasting approach which means that the reporting for training and testing considers every trading event under a rolling windows setting. Every stock was trained separately under two different feature sets, named Simple and Extended. Simple protocol refers to forecasting based on the best LOB level information (i.e., best ask and best bid together with their corresponding trading volumes). The Extended set refers to a larger, in terms of number of features (see \hyperref[tab:feat]{Table \ref{tab:feat}}), set that operates under the same experimental routines similar to Simple. Both Simple and Extended compete for the lowest MSE and RMSE scores and we present the results in Appendix in tables \hyperref[tab:simplemse]{Table \ref{tab:simplemse}} - \hyperref[tab:extermse]{Table \ref{tab:extermse}}. Every table contains 20 stocks over the period of three months (i.e., from September to November 2022). We provide with bold letters the best test performance (i.e., MDI Test against GD Test) method per stock per month. Every table refers to a specific input set (i.e., Simple or Extended) and a specific metric (i.e., MSE or RMSE). 

The results suggest that every stock exhibits different performance behaviour with respect to MDI and GD. Despite the monthly reporting, it should be stated that the performance profile is constantly changing between these two methods on a nanosecond level. All the stocks exhibited a constant alternation between two and three number of clusters within MDI and a rapid rate of change between the two feature importance methods - on average, there was a regime change from MDI to GD every 10 trading events. Depending on the input information (i.e., Simple or Extended feature set) stocks behaved differently. For instance, the performance behavior of MSFT under the Simple data set was worse compared to the Extended input. Also MSFT under the Simple set performed better based on the GD algorithm compared to the Extended data set that selected the MDI. Both feature importance methods offered low RMSE scores in several cases. For instance, GOOGL under the GD approach and based on the Extended feature set achieved the lowest RMSE score compared to the rest experimental scenarios. To highlight the significance of the fully autonomous protocol we also consider a normalized RMSE with respect to LOB's mid-price, named relative RMSE (RRMSE) that is based on the ratio between RMSE and the current mid-price. RRMSE enables us to compare directly the four proposed experimental settings (i.e., two feature importance methods across two different input sets) per stock. As outlined in \hyperref[tab:cv]{Table \ref{tab:cv}} our findings indicate that the MDI method, under the Simple feature set, achieved the lowest RRMSE for 36 out of 60 cases (i.e., across three months per stock). The primary purpose of these two metrics (i.e., RMSE and RRMSE) is to identify a stock-specific forecasting method capable to achieve low errors relative to the fluctuations of the forecasting objective, which in our study,is the level of the mid-price.  

\subsection{Limitations and Future Research}
\noindent Despite the efficacy of the proposed fully automated approach, our study does have certain limitations. Firstly, the study can be considered a narrow AI approach, which means that the experimental protocol developed is specifically tailored to this particular objective - the prediction of the Limit Order Book's (LOB's) mid-price. Additionally, our method utilizes certain specific features that can be easily engineered from the existing best LOB price level. In future work, it would be beneficial to employ more sophisticated, hand-crafted, and fully automated features. A further limitation is the lack of an extensive benchmark modeling framework to provide a more rigorous challenge to the existing RBFNN topology. Moreover, we made the assumption of isotropic clusters within the k-means algorithm, which presumes that every cluster exhibits constant variance. This assumption might not always hold true. Finally, we believe that the length of the utilized datasets should be extended in future studies to improve the robustness of the results.

\section{Conclusion}\label{section:Con}
\noindent Online forecasting in the HFT universe requires swift and fully autonomous mechanisms that can effectively utilize all available information. To the best of our knowledge, this study is the first to automate the process of defining the number of clusters within the k-means algorithm and RBFNN for tick-by-tick LOB's mid-price forecasting. We developed a competitive framework that consistently challenges the proposed MDI by converting GD into a feature importance method. We used data from 20 US mega cap stocks from Refinitiv, with nanosecond time resolution. Our findings suggest that an autonomous approach to clustering and feature importance presents challenges, but also provides several benefits for the machine learning-based trader.

\section*{Acknowledgments}
\noindent The authors would like to thank ECDF Linux Compute Cluster (Eddie) for the generous computational resources.

\appendices
\section{}
\label{sec:FirstAppendix}

\begin{table*}[!htbp]
\centering
\captionsetup{width=.90\textwidth}
\caption{MSE scores for the Simple experimental protocol.}
\scalebox{1.05}{
\begin{tabular}{crcccc}
\toprule
\textbf{Stock} & \textbf{Month} & \textbf{MDI Train} & \textbf{MDI Test} & \textbf{GD Train} & \textbf{GD Test} \\
\hline
MSFT & September & 4.064E-02 & 9.900E+02 & 1.032E-01 & \textbf{1.281E+02} \\
     & October   & 2.801E-02 & 4.723E+00 & 4.908E-02 & \textbf{1.972E-01} \\
     & November  & 6.668E-03 & \textbf{8.473E-02} & 8.743E-03 & 7.697E-01 \\
\hline
GOOGL & September & 6.155E-03 & 8.593E+01 & 6.758E-02 & \textbf{1.874E+00} \\
     & October    & 2.299E-02 & \textbf{7.808E-01} & 5.921E-01 & 4.584E+00 \\
     & November   & 1.662E-03 & \textbf{2.434E-02} & 1.148E-01 & 7.251E-01 \\
\hline
AMZN & September & 3.193E-02 & 1.256E+02 & 7.742E-01 & \textbf{5.903E+01} \\
     & October   & 3.961E-03 & \textbf{6.564E+00} & 1.785E-01 & 1.026E+01 \\
     & November  & 2.090E-03 & 5.582E-03 & 1.751E-04 & \textbf{1.021E-04} \\
\hline
BRK & September & 7.681E+07 & \textbf{7.670E+07} & 1.212E+08 & 1.208E+08 \\
    & October   & 4.498E+06 & \textbf{4.470E+06} & 7.762E+06 & 7.761E+06 \\
    & November  & 1.916E+08 & 1.884E+08 & 5.938E+04 & \textbf{5.817E+04} \\
\hline
NVDA & September & 3.864E-02 & 9.275E-01 & 5.663E-01 & \textbf{6.960E-01} \\
     & October   & 1.600E-01 & \textbf{1.466E-01} & 1.633E-01 & 3.076E-01 \\
     & November  & 2.084E-02 & 3.204E+03 & 1.226E-01 & \textbf{4.535E+01} \\
\hline
META & September & 3.098E-02 & 5.943E+02 & 2.003E-01 & \textbf{2.774E+01} \\
     & October   & 6.247E-02 & 6.773E+03 & 7.852E-02 & \textbf{1.809E+01} \\
     & November  & 1.275E-02 & 1.452E+02 & 1.736E-03 & \textbf{4.452E-03} \\
\hline
JNJ & September  & 1.941E-04 & \textbf{2.307E-04} & 4.555E-03 & 4.565E-03 \\
    & October    & 2.751E-02 & 2.762E-02 & 2.537E-04 & \textbf{3.322E-04} \\
    & November   & 4.311E-04 & \textbf{4.288E-04} & 2.918E-03 & 2.963E-03 \\
\hline
BAC & September  & 3.692E-03 & \textbf{9.912E-01} & 8.080E-01 & 3.976E+00 \\
    & October    & 4.142E-06 & \textbf{5.750E-03} & 9.853E-01 & 7.305E+00 \\
    & November   & 5.948E-06 & \textbf{2.309E-04} & 4.077E-01 & 1.248E+00 \\
\hline
HD & September & 5.974E-03 & 5.997E-03 & 3.898E-04 & \textbf{3.246E-04} \\
   & October & 3.376E-02 & \textbf{3.369E-02} & 1.637E-01 & 1.636E-01 \\
   & November & 3.075E-05 & \textbf{5.686E-04} & 2.544E-03 & 1.814E-03 \\
\hline
JPM & September & 1.255E-02 & \textbf{1.467E-02} & 3.386E-02 & 3.238E-02 \\
    & October & 1.157E-04 & \textbf{1.889E+00} & 5.889E-02 & 2.294E+01 \\
    & November & 3.605E-04 & 3.338E-04 & 1.087E-05 & \textbf{4.781E-05} \\
\hline
KO & September & 9.889E-06 & \textbf{6.942E-05} & 7.965E-04 & 6.968E-04 \\
   & October & 2.232E-05 & \textbf{2.340E-05} & 1.095E-03 & 4.486E-03 \\
   & November & 1.258E-05 & \textbf{4.406E-04} & 3.510E-03 & 1.711E-01 \\
\hline
LLY & September & 9.563E-05 & 1.253E-03 & 1.163E-03 & \textbf{8.080E-04} \\
    & October & 2.322E-03 & \textbf{4.930E-03} & 2.910E-01 & 2.814E-01 \\
    & November & 6.970E-04 & 7.249E-04 & 4.762E-04 & \textbf{4.654E-04} \\
\hline
MRK & September & 3.752E-05 & \textbf{2.931E-04} & 6.679E-05 & 8.882E-04 \\
    & October & 5.128E-02 & 4.648E-02 & 5.049E-04 & \textbf{1.510E-03} \\
    & November & 7.886E-05 & 1.545E-03 & 1.682E-04 & \textbf{4.444E-04} \\
\hline
NVO & September & 5.795E-05 & 1.212E-04 & 1.972E-07 & \textbf{4.189E-05} \\
    & October & 1.154E-06 & \textbf{2.686E-06} & 3.597E-06 & 8.483E-06 \\
    & November & 2.018E-04 & 2.937E-04 & 1.973E-07 & \textbf{1.320E-04} \\
\hline
ORCL & September & 1.532E-05 & \textbf{4.863E-03} & 2.089E-02 & 2.432E-02 \\
     & October & 6.832E-05 & \textbf{2.898E-04} & 1.913E-02 & 7.202E-02 \\
     & November & 1.035E-04 & \textbf{1.086E-04} & 2.315E-04 & 3.356E-03 \\
\hline
PEP & September & 2.441E-01 & 3.976E+02 & 1.038E+00 & \textbf{2.713E+00} \\
    & October & 9.905E-03 & \textbf{2.826E+01} & 4.468E-01 & 1.138E+02 \\
    & November & 5.510E-01 & \textbf{7.434E+00} & 1.625E+00 & 1.338E+01 \\
\hline
PG & September & 1.002E-04 & 4.258E-04 & 1.624E-05 & \textbf{3.598E-05} \\
   & October & 4.938E-04 & 4.801E-04 & 4.195E-05 & \textbf{4.502E-05} \\
   & November & 8.169E-04 & 8.261E-04 & 1.401E-06 & \textbf{3.304E-05} \\
\hline
UNH & September & 3.609E-02 & \textbf{3.613E-02} & 1.679E+02 & 1.678E+02 \\
    & October & 5.853E-02 & \textbf{5.653E-02} & 3.059E+00 & 3.034E+00 \\
    & November & 3.725E-03 & \textbf{3.839E-03} & 1.017E-02 & 1.465E-02 \\
\hline
VISA & September & 2.280E-04 & \textbf{2.930E-04} & 2.141E-03 & 2.387E-03 \\
     & October & 1.030E-02 & 2.024E-02 & 5.632E-04 & \textbf{1.289E-02} \\
     & November & 3.115E-02 & 3.085E-02 & 1.430E-03 & \textbf{1.477E-03} \\
\hline
WMT & September & 1.842E-04 & 2.128E-04 & 8.221E-05 & \textbf{1.605E-04} \\
    & October & 1.688E-04 & 2.868E-04 & 2.371E-05 & \textbf{1.570E-04} \\
    & November & 1.652E-04 & 5.494E-04 & 4.375E-04 & \textbf{4.090E-04} \\
\bottomrule
\end{tabular}}
\label{tab:simplemse}
\end{table*}

\begin{table*}[!htbp]
\centering
\captionsetup{width=.90\textwidth}
\caption{MSE scores for the Extended experimental protocol.}
\scalebox{1.05}{
\begin{tabular}{crcccc}
\toprule
\textbf{Stock} & \textbf{Month} & \textbf{MDI Train} & \textbf{MDI Test} & \textbf{GD Train} & \textbf{GD Test} \\
\hline
MSFT & September & 7.571E+00 & \textbf{3.168E+01} & 4.326E+01 & 4.318E+01 \\
     & October & 4.213E+00 & \textbf{5.496E+00} & 7.624E+01 & 7.620E+01 \\
     & November & 1.096E+01 & 6.046E+00 & 3.673E-03 & \textbf{1.243E-01} \\
\hline
GOOGL & September & 1.384E+01 & 5.530E+00 & 1.844E-03 & \textbf{2.540E-02} \\
     & October & 8.134E-01 & 2.426E+01 & 7.406E-03 & \textbf{1.406E-02} \\
     & November & 2.881E+00 & 2.423E+00 & 1.963E-03 & \textbf{1.675E-03} \\
\hline
AMZN & September & 6.304E+00 & 1.019E+01 & 2.707E-01 & \textbf{3.729E-01} \\
     & October & 1.753E+00 & 2.194E+00 & 4.789E-01 & \textbf{5.057E-01} \\
     & November & 1.293E+00 & 1.297E+00 & 1.159E-02 & \textbf{1.401E-02} \\
\hline
BRK & September & 3.497E+07 & 3.466E+07 & 5.861E+06 & \textbf{5.701E+06} \\
    & October & 2.965E+08 & 2.924E+08 & 1.221E+08 & \textbf{1.197E+08} \\
    & November & 2.825E+08 & 2.584E+08 & 1.184E+08 & \textbf{1.086E+08} \\
\hline
NVDA & September & 1.754E-01 & 3.359E-01 & 5.952E-03 & \textbf{2.362E-02} \\
     & October & 1.088E+02 & 1.090E+02 & 1.421E-01 & \textbf{1.902E-01} \\
     & November & 6.868E-01 & 1.046E+02 & 2.127E-03 & \textbf{2.393E-02} \\
\hline
META & September & 1.214E-01 & 2.687E+00 & 2.764E-03 & \textbf{8.930E-03} \\
     & October & 6.670E-02 & 1.272E+03 & 5.459E-03 & \textbf{4.523E+00} \\
     & November & 5.232E+02 & 5.531E+02 & 5.168E-02 & \textbf{5.505E-01} \\
\hline
JNJ & September & 1.886E+02 & 1.879E+02 & 7.664E-03 & \textbf{8.395E-02} \\
    & October & 1.220E+01 & 1.749E+02 & 2.615E-03 & \textbf{3.423E-01} \\
    & November & 4.984E+02 & 4.940E+02 & 8.931E-04 & \textbf{4.912E-02} \\
\hline
BAC & September & 1.565E-07 & \textbf{1.650E-02} & 1.269E-01 & 1.738E-01 \\
    & October & 3.564E-06 & 1.035E+01 & 1.737E-02 & \textbf{1.789E-01} \\
    & November & 4.357E-07 & 4.688E-01 & 2.260E-03 & \textbf{1.757E-02} \\
\hline
HD & September & 3.533E+00 & 1.642E+02 & 1.145E-01 & \textbf{7.436E-03} \\
   & October & 2.423E-01 & 2.929E-01 & 3.140E-02 & \textbf{5.118E-02} \\
   & November & 6.361E-01 & 7.064E-01 & 2.558E-02 & \textbf{2.043E-01} \\
\hline
JPM & September & 2.114E+00 & \textbf{1.685E+00} & 1.772E+00 & 1.844E+00 \\
    & October & 1.623E+00 & 4.416E+00 & 1.353E-02 & \textbf{4.690E-02} \\
    & November & 8.077E+00 & 4.971E+01 & 7.347E-01 & \textbf{7.326E-01} \\
\hline
KO & September & 2.298E-04 & \textbf{3.866E+00} & 7.605E-02 & 2.546E+01 \\
   & October & 6.014E-05 & 1.903E+01 & 2.228E-03 & \textbf{3.472E-02} \\
   & November & 4.111E-05 & \textbf{6.374E-03} & 1.950E-03 & 5.919E-02 \\
\hline
LLY & September & 4.994E+01 & 4.966E+01 & 4.274E-03 & \textbf{8.325E-01} \\
    & October & 2.655E+02 & 2.487E+02 & 1.461E-02 & \textbf{7.177E+00} \\
    & November & 9.689E+00 & \textbf{9.110E+00} & 1.903E-04 & 1.279E+01 \\
\hline
MRK & September & 1.282E+00 & 1.273E+00 & 1.586E-02 & \textbf{2.234E-02} \\
    & October & 5.880E-01 & 2.641E-01 & 2.678E-03 & \textbf{4.601E-02} \\
    & November & 8.285E-02 & 4.956E+00 & 8.898E-04 & \textbf{1.036E-01} \\
\hline
NVO & September & 1.702E+00 & 1.264E+00 & 1.169E-02 & \textbf{1.763E-01} \\
    & October & 2.997E-01 & 2.938E-01 & 1.391E-04 & \textbf{1.582E-01} \\
    & November & 4.161E-01 & 6.163E-01 & 1.013E-04 & \textbf{1.243E-01} \\
\hline
ORCL & September & 3.209E+00 & 2.939E+01 & 8.546E-03 & \textbf{8.063E-02} \\
     & October & 3.049E-02 & 1.620E+01 & 3.691E-03 & \textbf{5.246E-02} \\
     & November & 2.197E-02 & 8.574E-02 & 6.022E-04 & \textbf{2.517E-02} \\
\hline
PEP & September & 7.846E-04 & \textbf{4.445E-01} & 5.668E-03 & 6.787E+00 \\
    & October & 2.865E-02 & 2.704E+01 & 3.086E-04 & \textbf{2.699E-04} \\
    & November & 9.896E-03 & \textbf{8.882E-03} & 2.575E-04 & 8.817E-01 \\
\hline
PG & September & 6.923E-02 & 8.616E-01 & 4.738E-03 & \textbf{1.092E-01} \\
   & October & 5.011E+00 & 4.183E+00 & 2.220E-03 & \textbf{3.142E-02} \\
   & November & 4.965E+00 & 1.612E+01 & 9.110E-03 & \textbf{3.765E-01} \\
\hline
UNH & September & 3.817E+01 & 3.775E+01 & 6.593E-01 & \textbf{6.598E-01} \\
    & October & 8.824E-01 & 9.665E-01 & 4.806E-02 & \textbf{8.800E-02} \\
    & November & 1.389E+00 & 1.398E+00 & 4.276E-03 & \textbf{4.401E-03} \\
\hline
VISA & September & 1.015E+01 & 1.485E+01 & 6.168E-04 & \textbf{1.078E-01} \\
     & October & 3.122E+01 & 1.548E+03 & 5.629E-04 & \textbf{2.598E-01} \\
     & November & 9.065E-02 & \textbf{7.932E-02} & 1.443E-01 & 4.912E-01 \\
\hline
WMT & September & 1.059E+00 & 8.102E-01 & 5.356E-03 & \textbf{9.036E-03} \\
    & October & 1.317E+01 & 3.035E+02 & 1.336E-03 & \textbf{1.182E-01} \\
    & November & 2.002E+03 & 1.988E+03 & 4.379E-04 & \textbf{2.932E-01} \\
\bottomrule
\end{tabular}}
\label{tab:extemse}
\end{table*}

\begin{table*}[!htbp]
\centering
\captionsetup{width=.90\textwidth}
\caption{RMSE scores for the Simple experimental protocol.}
\scalebox{1.05}{
\begin{tabular}{crcccc}
\toprule
\textbf{Stock} & \textbf{Month} & \textbf{MDI Train} & \textbf{MDI Test} & \textbf{GD Train} & \textbf{GD Test} \\
\hline
MSFT   & September & 2.016E-01 & 3.147E+01 & 3.212E-01 & \textbf{1.132E+01} \\
       & October & 1.674E-01 & 2.173E+00 & 2.215E-01 & \textbf{4.441E-01} \\
       & November & 8.166E-02 & \textbf{2.911E-01} & 9.350E-02 & 8.773E-01 \\
\hline
GOOGL   & September & 7.845E-02 & 9.270E+00 & 2.600E-01 & \textbf{1.369E+00} \\
       & October & 1.516E-01 & \textbf{8.836E-01} & 7.695E-01 & 2.141E+00 \\
       & November & 4.077E-02 & \textbf{1.560E-01} & 3.389E-01 & 8.515E-01 \\
\hline
AMZN   & September & 1.787E-01 & 1.121E+01 & 8.799E-01 & \textbf{7.683E+00} \\
       & October & 6.293E-02 & \textbf{2.562E+00} & 4.225E-01 & 3.204E+00 \\
       & November & 4.572E-02 & 7.471E-02 & 1.323E-02 & \textbf{1.011E-02} \\
\hline
BRK    & September & 8.764E+03 & \textbf{8.758E+03} & 1.101E+04 & 1.099E+04 \\
       & October & 2.121E+03 & \textbf{2.114E+03} & 2.786E+03 & 2.786E+03 \\
       & November & 1.384E+04 & 1.372E+04 & 2.437E+02 & \textbf{2.412E+02} \\
\hline
NVDA & September & 1.966E-01 & 9.630E-01 & 7.525E-01 & \textbf{8.343E-01} \\
     & October & 4.000E-01 & \textbf{3.829E-01} & 4.041E-01 & 5.546E-01 \\
     & November & 1.443E-01 & 5.661E+01 & 3.501E-01 & \textbf{6.734E+00} \\
\hline
META & September & 1.760E-01 & 2.438E+01 & 4.476E-01 & \textbf{5.267E+00} \\
     & October & 2.499E-01 & 8.230E+01 & 2.802E-01 & \textbf{4.253E+00} \\
     & November & 1.129E-01 & 1.205E+01 & 4.166E-02 & \textbf{6.673E-02} \\
\hline
JNJ & September & 1.393E-02 & \textbf{1.519E-02} & 6.749E-02 & 6.757E-02 \\
    & October & 1.659E-01 & 1.662E-01 & 1.593E-02 & \textbf{1.823E-02} \\
    & November & 2.076E-02 & \textbf{2.071E-02} & 5.402E-02 & 5.443E-02 \\
\hline
BAC & September & 6.076E-02 & \textbf{9.956E-01} & 8.989E-01 & 1.994E+00 \\
    & October & 2.035E-03 & \textbf{7.583E-02} & 9.926E-01 & 2.703E+00 \\
    & November & 2.439E-03 & \textbf{1.519E-02} & 6.385E-01 & 1.117E+00 \\
\hline
HD  & September & 7.729E-02 & 7.744E-02 & 1.974E-02 & \textbf{1.802E-02} \\
    & October & 1.837E-01 & \textbf{1.836E-01} & 4.046E-01 & 4.045E-01 \\
    & November & 5.546E-03 & \textbf{2.385E-02} & 5.044E-02 & 4.259E-02 \\
\hline
JPM & September & 1.120E-01 & \textbf{1.211E-01} & 1.840E-01 & 1.800E-01 \\
    & October & 1.076E-02 & \textbf{1.374E+00} & 2.427E-01 & 4.790E+00 \\
    & November & 1.899E-02 & 1.827E-02 & 3.297E-03 & \textbf{6.915E-03} \\
\hline
KO  & September & 3.145E-03 & \textbf{8.332E-03} & 2.822E-02 & 2.640E-02 \\
    & October & 4.725E-03 & \textbf{4.837E-03} & 3.309E-02 & 6.698E-02 \\
    & November & 3.547E-03 & \textbf{2.099E-02} & 5.925E-02 & 4.136E-01 \\
\hline
LLY & September & 9.779E-03 & 3.540E-02 & 3.411E-02 & \textbf{2.842E-02} \\
    & October & 4.819E-02 & \textbf{7.022E-02} & 5.394E-01 & 5.304E-01 \\
    & November & 2.640E-02 & 2.692E-02 & 2.182E-02 & \textbf{2.157E-02} \\
\hline
MRK & September & 6.125E-03 & \textbf{1.712E-02} & 8.172E-03 & 2.980E-02 \\
    & October & 2.265E-01 & 2.156E-01 & 2.247E-02 & \textbf{3.886E-02} \\
    & November & 8.880E-03 & 3.931E-02 & 1.297E-02 & \textbf{2.108E-02} \\
\hline
NVO & September & 7.612E-03 & 1.101E-02 & 4.441E-04 & \textbf{6.472E-03} \\
    & October & 1.074E-03 & \textbf{1.639E-03} & 1.897E-03 & 2.913E-03 \\
    & November & 1.420E-02 & 1.714E-02 & 4.442E-04 & \textbf{1.149E-02} \\
\hline
ORCL & September & 3.914E-03 & \textbf{6.973E-02} & 1.445E-01 & 1.560E-01 \\
     & October & 8.266E-03 & \textbf{1.702E-02} & 1.383E-01 & 2.684E-01 \\
     & November & 1.017E-02 & \textbf{1.042E-02} & 1.522E-02 & 5.793E-02 \\
\hline
PEP & September & 4.941E-01 & 1.994E+01 & 1.019E+00 & \textbf{1.647E+00} \\
    & October & 9.952E-02 & \textbf{5.316E+00} & 6.684E-01 & 1.067E+01 \\
    & November & 7.423E-01 & \textbf{2.727E+00} & 1.275E+00 & 3.658E+00 \\
\hline
PG & September & 1.001E-02 & 2.063E-02 & 4.030E-03 & \textbf{5.998E-03} \\
   & October & 2.222E-02 & 2.191E-02 & 6.477E-03 & \textbf{6.710E-03} \\
   & November & 2.858E-02 & 2.874E-02 & 1.184E-03 & \textbf{5.748E-03} \\
\hline
UNH & September & 1.900E-01 & \textbf{1.901E-01} & 1.296E+01 & 1.296E+01 \\
    & October & 2.419E-01 & \textbf{2.378E-01} & 1.749E+00 & 1.742E+00 \\
    & November & 6.103E-02 & \textbf{6.196E-02} & 1.008E-01 & 1.210E-01 \\
\hline
VISA & September & 1.510E-02 & \textbf{1.712E-02} & 4.627E-02 & 4.886E-02 \\
     & October & 1.015E-01 & 1.423E-01 & 2.373E-02 & \textbf{1.136E-01} \\
     & November & 1.765E-01 & 1.756E-01 & 3.781E-02 & \textbf{3.844E-02} \\
\hline
WMT & September & 1.357E-02 & 1.459E-02 & 9.067E-03 & \textbf{1.267E-02} \\
    & October & 1.299E-02 & 1.694E-02 & 4.869E-03 & \textbf{1.253E-02} \\
    & November & 1.285E-02 & 2.344E-02 & 2.092E-02 & \textbf{2.022E-02} \\
\bottomrule
\end{tabular}}
\label{tab:simple_rmse}
\end{table*}

\begin{table*}[!htbp]
\centering
\captionsetup{width=.90\textwidth}
\caption{RMSE scores for the Extended experimental protocol.}
\scalebox{1.05}{
\begin{tabular}{crcccc}
\toprule
\textbf{Stock} & \textbf{Month} & \textbf{MDI Train} & \textbf{MDI Test} & \textbf{GD Train} & \textbf{GD Test} \\
\hline
MSFT & September & 2.752E+00 & \textbf{5.629E+00} & 6.577E+00 & 6.571E+00 \\
     & October & 2.053E+00 & \textbf{2.344E+00} & 8.732E+00 & 8.729E+00 \\
     & November & 3.311E+00 & 2.459E+00 & 6.061E-02 & \textbf{3.525E-01} \\
\hline
GOOGL & September & 3.720E+00 & 2.352E+00 & 4.295E-02 & \textbf{1.594E-01} \\
     & October & 9.019E-01 & 4.926E+00 & 8.606E-02 & \textbf{1.186E-01} \\
     & November & 1.697E+00 & 1.557E+00 & 4.431E-02 & \textbf{4.092E-02} \\
\hline
AMZN & September & 2.511E+00 & 3.192E+00 & 5.203E-01 & \textbf{6.107E-01} \\
     & October & 1.324E+00 & 1.481E+00 & 6.920E-01 & \textbf{7.112E-01} \\
     & November & 1.137E+00 & 1.139E+00 & 1.077E-01 & \textbf{1.183E-01} \\
\hline
BRK & September & 5.914E+03 & 5.887E+03 & 2.421E+03 & \textbf{2.388E+03} \\
    & October & 1.722E+04 & 1.710E+04 & 1.105E+04 & \textbf{1.094E+04} \\
    & November & 1.681E+04 & 1.608E+04 & 1.088E+04 & \textbf{1.042E+04} \\
\hline
NVDA & September & 4.188E-01 & 5.796E-01 & 7.715E-02 & \textbf{1.537E-01} \\
     & October & 1.043E+01 & 1.044E+01 & 3.770E-01 & \textbf{4.361E-01} \\
     & November & 8.287E-01 & 1.023E+01 & 4.612E-02 & \textbf{1.547E-01} \\
\hline
META & September & 3.484E-01 & 1.639E+00 & 5.258E-02 & \textbf{9.450E-02} \\
     & October & 2.583E-01 & 3.566E+01 & 7.389E-02 & \textbf{2.127E+00} \\
     & November & 2.287E+01 & 2.352E+01 & 2.273E-01 & \textbf{7.420E-01} \\
\hline
JNJ & September & 1.373E+01 & 1.371E+01 & 8.755E-02 & \textbf{2.897E-01} \\
    & October & 3.492E+00 & 1.323E+01 & 5.114E-02 & \textbf{5.851E-01} \\
    & November & 2.232E+01 & 2.223E+01 & 2.989E-02 & \textbf{2.216E-01} \\
\hline
BAC & September & 3.956E-04 & \textbf{1.285E-01} & 3.562E-01 & 4.169E-01 \\
    & October & 1.888E-03 & 3.217E+00 & 1.318E-01 & \textbf{4.230E-01} \\
    & November & 6.600E-04 & 6.847E-01 & 4.754E-02 & \textbf{1.325E-01} \\
\hline
HD  & September & 1.880E+00 & 1.281E+01 & 3.384E-01 & \textbf{8.623E-02} \\
    & October & 4.922E-01 & 5.412E-01 & 1.772E-01 & \textbf{2.262E-01} \\
    & November & 7.976E-01 & 8.405E-01 & 1.599E-01 & \textbf{4.520E-01} \\
\hline
JPM & September & 1.454E+00 & \textbf{1.298E+00} & 1.331E+00 & 1.358E+00 \\
    & October & 1.274E+00 & 2.101E+00 & 1.163E-01 & \textbf{2.166E-01} \\
    & November & 2.842E+00 & 7.050E+00 & 8.571E-01 & \textbf{8.559E-01} \\
\hline
KO  & September & 1.516E-02 & \textbf{1.966E+00} & 2.758E-01 & 5.045E+00 \\
    & October & 7.755E-03 & 4.362E+00 & 4.720E-02 & \textbf{1.863E-01} \\
    & November & 6.411E-03 & \textbf{7.984E-02} & 4.416E-02 & 2.433E-01 \\
\hline
LLY & September & 7.067E+00 & 7.047E+00 & 6.538E-02 & \textbf{9.124E-01} \\
    & October & 1.629E+01 & 1.577E+01 & 1.209E-01 & \textbf{2.679E+00} \\
    & November & 3.113E+00 & \textbf{3.018E+00} & 1.379E-02 & 3.576E+00 \\
\hline
MRK & September & 1.132E+00 & 1.128E+00 & 1.259E-01 & \textbf{1.495E-01} \\
    & October & 7.668E-01 & 5.139E-01 & 5.175E-02 & \textbf{2.145E-01} \\
    & November & 2.878E-01 & 2.226E+00 & 2.983E-02 & \textbf{3.218E-01} \\
\hline
NVO & September & 1.305E+00 & 1.124E+00 & 1.081E-01 & \textbf{4.198E-01} \\
    & October & 5.475E-01 & 5.421E-01 & 1.179E-02 & \textbf{3.977E-01} \\
    & November & 6.451E-01 & 7.851E-01 & 1.006E-02 & \textbf{3.526E-01} \\
\hline
ORCL & September & 1.791E+00 & 5.421E+00 & 9.244E-02 & \textbf{2.840E-01} \\
     & October & 1.746E-01 & 4.024E+00 & 6.076E-02 & \textbf{2.290E-01} \\
     & November & 1.482E-01 & 2.928E-01 & 2.454E-02 & \textbf{1.587E-01} \\
\hline
PEP & September & 2.801E-02 & \textbf{6.667E-01} & 7.529E-02 & 2.605E+00 \\
    & October & 1.693E-01 & 5.200E+00 & 1.757E-02 & \textbf{1.643E-02} \\
    & November & 9.948E-02 & \textbf{9.424E-02} & 1.605E-02 & 9.390E-01 \\
\hline
PG  & September & 2.631E-01 & 9.282E-01 & 6.883E-02 & \textbf{3.304E-01} \\
    & October & 2.238E+00 & 2.045E+00 & 4.712E-02 & \textbf{1.773E-01} \\
    & November & 2.228E+00 & 4.015E+00 & 9.545E-02 & \textbf{6.136E-01} \\
\hline
UNH & September & 6.178E+00 & 6.144E+00 & 8.120E-01 & \textbf{8.123E-01} \\
    & October & 9.394E-01 & 9.831E-01 & 2.192E-01 & \textbf{2.966E-01} \\
    & November & 1.179E+00 & 1.183E+00 & 6.539E-02 & \textbf{6.634E-02} \\
\hline
VISA & September & 3.185E+00 & 3.853E+00 & 2.484E-02 & \textbf{3.284E-01} \\
     & October & 5.587E+00 & 3.934E+01 & 2.373E-02 & \textbf{5.097E-01} \\
     & November & 3.011E-01 & \textbf{2.816E-01} & 3.798E-01 & 7.009E-01 \\
\hline
WMT & September & 1.029E+00 & 9.001E-01 & 7.319E-02 & \textbf{9.506E-02} \\
    & October & 3.629E+00 & 1.742E+01 & 3.655E-02 & \textbf{3.438E-01} \\
    & November & 4.474E+01 & 4.459E+01 & 2.093E-02 & \textbf{5.415E-01} \\
\bottomrule
\end{tabular}}
\label{tab:extermse}
\end{table*}

\begin{table*}[!htbp]
\centering
\captionsetup{width=.90\textwidth}
\caption{RRMSE scores for the Simple (Sim) and Extended (Exte) test experimental protocols.}
\scalebox{1.00}{
\begin{tabular}{crcccc}
\toprule
\textbf{Stock} & \textbf{Month} & \textbf{RRMSE MDI Sim} & \textbf{RRMSE GD Sim} & \textbf{RRMSE MDI Exte} & \textbf{RRMSE GD Exte} \\ 
\hline
MSFT & September & \textbf{9.300E-02} & 1.451E+01 & 1.482E-01 & 5.220E+00 \\
     & October & \textbf{7.782E-02} & 1.010E+00 & 1.030E-01 & 2.065E-01 \\
     & November & \textbf{3.970E-02} & 1.415E-01 & 4.547E-02 & 4.266E-01 \\
\hline
GOOG & September & \textbf{2.586E-02} & 3.055E+00 & 8.568E-02 & 4.512E-01 \\
     & October & \textbf{4.454E-02} & 2.596E-01 & 2.260E-01 & 6.289E-01 \\
     & November & \textbf{9.948E-03} & 3.806E-02 & 8.268E-02 & 2.078E-01 \\
\hline
AMZN & September & \textbf{5.807E-02} & 3.642E+00 & 2.860E-01 & 2.497E+00 \\
     & October & \textbf{1.777E-02} & 7.233E-01 & 1.193E-01 & 9.045E-01 \\
     & November & 9.286E-03 & 1.518E-02 & 2.688E-03 & \textbf{2.053E-03} \\
\hline
BRK & September & 8.010E+03 & \textbf{8.005E+03} & 1.006E+04 & 1.005E+04 \\
    & October & 1.960E+03 & \textbf{1.954E+03} & 2.575E+03 & 2.574E+03 \\
    & November & 1.194E+04 & 1.184E+04 & 2.102E+02 & \textbf{2.080E+02} \\
\hline
NVDA & September & \textbf{1.273E-01} & 6.236E-01 & 4.873E-01 & 5.402E-01 \\
     & October & 1.386E-01 & \textbf{1.326E-01} & 1.400E-01 & 1.922E-01 \\
     & November & \textbf{5.574E-02} & 2.186E+01 & 1.352E-01 & 2.600E+00 \\
\hline
META & September & \textbf{5.704E-02} & 7.901E+00 & 1.451E-01 & 1.707E+00 \\
     & October & \textbf{5.884E-02} & 1.937E+01 & 6.597E-02 & 1.001E+00 \\
     & November & 2.619E-02 & 2.795E+00 & \textbf{9.663E-03} & 1.548E-02 \\
\hline
JNJ & September & \textbf{5.331E-03} & 5.812E-03 & 2.582E-02 & 2.585E-02 \\
    & October & 6.460E-02 & 6.473E-02 & \textbf{6.204E-03} & 7.099E-03 \\
    & November & 8.765E-03 & \textbf{8.742E-03} & 2.281E-02 & 2.298E-02 \\
\hline
BAC & September & \textbf{5.560E-04} & 9.110E-03 & 8.225E-03 & 1.825E-02 \\
    & October & \textbf{2.583E-05} & 9.623E-04 & 1.260E-02 & 3.430E-02 \\
    & November & \textbf{5.044E-05} & 3.143E-04 & 1.321E-02 & 2.310E-02 \\
\hline
HD & September & 5.557E-02 & 5.568E-02 & 1.419E-02 & \textbf{1.295E-02} \\
   & October & 1.319E-01 & \textbf{1.317E-01} & 2.904E-01 & 2.903E-01 \\
   & November & \textbf{3.236E-03} & 1.391E-02 & 2.943E-02 & 2.485E-02 \\
\hline
JPM & September & 3.882E-02 & \textbf{4.197E-02} & 6.378E-02 & 6.237E-02 \\
    & October & \textbf{3.638E-03} & 4.648E-01 & 8.207E-02 & 1.620E+00 \\
    & November & 7.185E-03 & 6.913E-03 & \textbf{1.247E-03} & 2.616E-03 \\
\hline
KO & September & \textbf{3.029E-04} & 8.025E-04 & 2.718E-03 & 2.543E-03 \\
   & October & \textbf{5.216E-04} & 5.340E-04 & 3.653E-03 & 7.394E-03 \\
   & November & \textbf{4.020E-04} & 2.379E-03 & 6.715E-03 & 4.688E-02 \\
\hline
LLY & September & \textbf{6.532E-03} & 2.365E-02 & 2.278E-02 & 1.899E-02 \\
    & October & \textbf{3.405E-02} & 4.961E-02 & 3.811E-01 & 3.748E-01 \\
    & November & 1.549E-02 & 1.580E-02 & 1.280E-02 & \textbf{1.266E-02} \\
\hline
MRK & September & \textbf{1.754E-03} & 4.902E-03 & 2.340E-03 & 8.532E-03 \\
    & October & 6.431E-02 & 6.123E-02 & \textbf{6.382E-03} & 1.104E-02 \\
    & November & \textbf{2.560E-03} & 1.133E-02 & 3.739E-03 & 6.077E-03 \\
\hline
NVO & September & 3.939E-03 & 5.698E-03 & \textbf{2.298E-04} & 3.349E-03 \\
    & October & \textbf{4.664E-04} & 7.116E-04 & 8.235E-04 & 1.265E-03 \\
    & November & 6.358E-03 & 7.671E-03 & \textbf{1.988E-04} & 5.143E-03 \\
\hline
ORCL & September & \textbf{8.691E-04} & 1.548E-02 & 3.210E-02 & 3.463E-02 \\
     & October & \textbf{1.854E-03} & 3.818E-03 & 3.102E-02 & 6.019E-02 \\
     & November & 2.976E-03 & 3.047E-03 & \textbf{4.450E-03} & 1.694E-02 \\
\hline
PEP & September & \textbf{1.503E-01} & 6.066E+00 & 3.099E-01 & 5.011E-01 \\
    & October & \textbf{2.759E-02} & 1.474E+00 & 1.853E-01 & 2.957E+00 \\
    & November & \textbf{3.421E-01} & 1.256E+00 & 5.874E-01 & 1.686E+00 \\
\hline
PG & September & 3.194E-03 & 6.583E-03 & \textbf{1.286E-03} & 1.914E-03 \\
   & October & 9.676E-03 & 9.541E-03 & \textbf{2.820E-03} & 2.922E-03 \\
   & November & 1.016E-02 & 1.022E-02 & 4.209E-04 & \textbf{2.044E-03} \\
\hline
UNH & September & \textbf{1.468E-01} & 1.469E-01 & 1.002E+01 & 1.001E+01 \\
    & October & 1.909E-01 & \textbf{1.876E-01} & 1.380E+00 & 1.375E+00 \\
    & November & \textbf{4.602E-02} & 4.672E-02 & 7.603E-02 & 9.127E-02 \\
\hline
VISA & September & \textbf{6.715E-03} & 7.612E-03 & 2.058E-02 & 2.173E-02 \\
     & October & 4.298E-02 & 6.024E-02 & \textbf{1.005E-02} & 4.809E-02 \\
     & November & 9.817E-02 & 9.769E-02 & \textbf{2.103E-02} & 2.138E-02 \\
\hline
WMT & September & 6.000E-03 & 6.450E-03 & \textbf{4.009E-03} & 5.601E-03 \\
    & October & 4.450E-03 & 5.801E-03 & \textbf{1.668E-03} & 4.292E-03 \\
    & November & \textbf{4.114E-03} & 7.503E-03 & 6.695E-03 & 6.473E-03 \\
\bottomrule
\end{tabular}}
\label{tab:cv}
\end{table*}

\end{document}